\documentclass[10pt, fleqn]{article}
\usepackage[utf8]{inputenc}
\DeclareUnicodeCharacter{2212}{$-$}
\usepackage{graphicx}
\usepackage{multirow}
\usepackage[table,xcdraw]{xcolor}
\usepackage[fleqn]{amsmath}
\usepackage[version=4]{mhchem}
\usepackage{siunitx}
\usepackage{longtable,tabularx}
\usepackage{nccmath}
\usepackage{amsmath}
\usepackage[superscript]{cite}
\usepackage[margin=1in]{geometry}
\usepackage[toc,page]{appendix}
\usepackage{setspace}
\usepackage{amssymb}
\usepackage{tabu}
\usepackage{listings}
\usepackage{afterpage}
\usepackage{titlesec}
\usepackage[labelfont=bf]{caption}
\usepackage{fancyhdr}
\usepackage{pgf}
\usepackage{float}
\usepackage{subcaption}
\usepackage{cleveref}
\usepackage{textcomp}
\usepackage{pict2e}
\usepackage{wasysym}
\usepackage{multicol}
\usepackage{lipsum}
\usepackage{indentfirst}
\newenvironment{Figure}
  {\par\medskip\noindent\minipage{\linewidth}}
  {\endminipage\par\medskip}

\titleformat{\section}
{\normalfont\normalsize\bfseries}{\thesection}{0.5em}{}
\titleformat{\subsection}
{\normalfont\normalsize\bfseries\em}{\thesubsection}{0.5em}{}
\titleformat{\subsubsection}
{\normalfont\normalsize\bfseries\em}{\thesubsubsection}{0.5em}{}
\begin{document}

\begin{flushright}
\Large 

\textbf{[SSC21-X-03]}
\end{flushright}
\begin{centering}      
\large 

\textbf{5G NB-IoT via low density LEO Constellations}\\
\vspace{0.5cm}
\normalsize 

{René Brandborg Sørensen}, Henrik Krogh Møller, Per Koch\\
{GateHouse SatCom}\\
{Strømmen 6, 9400 Nørresundby, Denmark}; \\
+4520994382, {rbs@gatehouse.com}

\vspace{0.5cm}
\centerline{\textbf{ABSTRACT}}
\vspace{0.3cm}
\end{centering}

5G NB-IoT is seen as a key technology for providing truly ubiquitous, global 5G coverage (1.000.000 devices/km2) for machine type communications in the internet of things. A non-terrestrial network (NTN) variant of NB-IoT is being standardized in the 3GPP, which along with inexpensive and non-complex chip-sets enables the production of competitively priced IoT devices with truly global coverage.
NB-IoT allows for narrowband single carrier transmissions in the uplink, which improves the uplink link-budget by as much as $16.8$~dB over the $180$~[kHz] downlink. This allows for a long range sufficient for ground to low earth orbit (LEO) communication without the need for complex and expensive antennas in the IoT devices.
In this paper the feasibility of 5G NB-IoT in the context of low-density constellations of small-satellites carrying base-stations in LEO is analyzed and required adaptations to NB-IoT are discussed.

\begin{multicols*}{2}

\section*{INTRODUCTION}
NB-IoT is a cellular protocol intended for service of massive number of devices in the internet of things (IoT). NB-IoT is a stripped down version of LTE, which removes much complexity from user equipment (UE) and utilizes narrow-band (NB) transmissions to achieve a better link budget. NB-IoT fulfills the requirement of connection density of 1.000.000 [devices/$\mathrm{km}^2$] for 5G, which can be supported by $10\times$ $200$~[kHz] carriers, i.e. $2$~[MHz] bandwidth.\cite{ciot} The 3rd Generation Partnership Project (3GPP) concluded a study on the support for non-terrestrial networks (NTN) for New Radio (NR) in 3GPP Release 16.\cite{TR38.811,TR38.821} A NTN version of NB-IoT is being pursued as a part of 3GPP Release 17 for 2022.\cite{TR36.763} 

IoT devices are typically inexpensive, low-complexity battery-powered devices with required life-times of several years in order to be economically viable. These devices are a stark contrast to the ordinary ground-stations typically used in satellite communication that may provide up to $100$ [W] of transmission power.\cite{7964683} The highest power class in NB-IoT is power class 3 (PC3) - a transmission power of $23$ [dBm] equivalent to $200$ [mW].

 NB-IoT is a viable candidate for providing uplink connectivity for power-constrained IoT devices in remote areas in the NTN scenario. NB-IoT in GEO is challenged by delays and propagation losses\cite{9268829} while NB-IoT in LEO is challenged by Doppler. NB-IoT utilizes narrow-band transmissions to achieve significant gains in the link-budget and the signal loss due to propagation is lowest in LEO where the distance between the UE and satellite will be smaller than in MEO and GEO. In general, MEO and GEO satellites benefit from being larger than LEO satellites, which typically have to be smaller and less expensive because of the air-drag and lower life-times in LEO. 



This paper investigates the viability of NTN NB-IoT in LEO primarily at a link-level with some system considerations for small satellites in low density constellations.

The paper is structured as follows: First, a system model is introduced presenting parameters for the satellites and IoT devices within the NTN. 
Then the link between the satellite and IoT device is analysed in terms of geometry, Doppler, propagation time and link budget before a discussion of the adaptations required in NB-IoT to support the LEO NTN case. Finally, the results presented throughout are discussed and concluding remarks are made.

\begin{Figure}
    \centering
    \includegraphics[width = 1\textwidth]{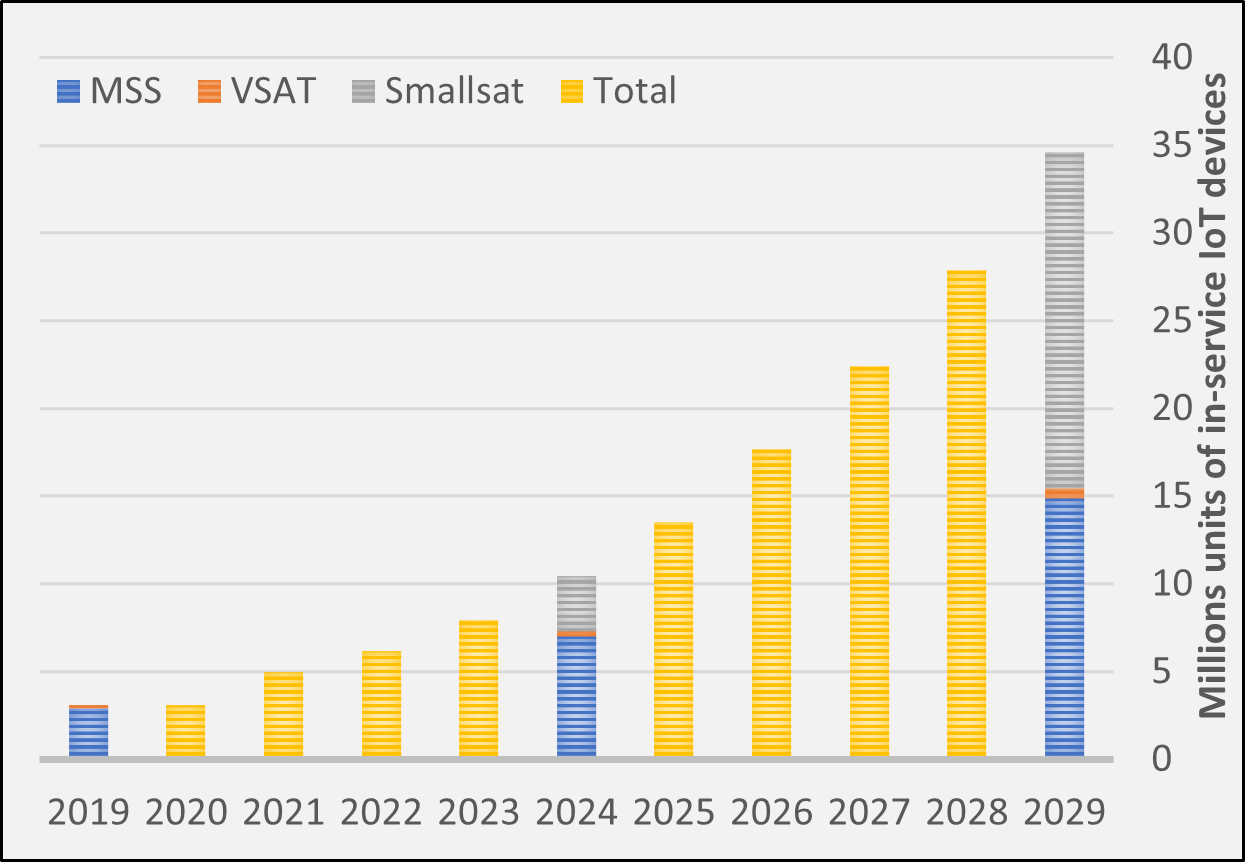}
    \captionof{figure}{Growth in IoT via satellite.}
    \label{fig:market}
\end{Figure}

\section*{SYSTEM MODEL}
In this section system level considerations for the LEO NTN NB-IoT use-case are presented. 

\subsection*{Scenario}
Consider the earth to be a sphere with radius $r_e  = 6357$~[km], rotating at a velocity $V_e  = 460$~[m/s] at the surface. The mass of earth is $M_e= 5.972\cdot10^{24}$~[kg].
A satellite orbits the earth in a Kepler orbit at a height $h_0=600$~[km] from the surface of the earth. The orbit is around Earths orbital axis, but in the opposite direction of Earths spin in order to achieve worst case Doppler. The satellite velocity is $v_{sat}  \approx \sqrt{(G\cdot M_e/(r_e+h_0 ))} = 7.57$~[km/s] .  

\subsection*{NB-IoT}
NB-IoT utilizes 180 kHz carriers, which may be deployed on a single LTE or NR carrier, within guard band or as a standalone technology. Here, NB-IoT is deployed as its standalone version and spectrum is allocated for it within the S-band (2GHz).
NB-IoT base-stations (eNB)s transmit in the entire $180$~[kHZ] in the downlink (DL) utilizing OFDM. The uplink (UL) is allocated spectrum on a separate carrier where SC-FDMA allow UEs to transmit, potentially simultaneously, in narrower bandwidths: $3.75$~[kHZ], $15$~[kHZ], $45$~[kHZ], $90$~[kHZ] and $180$~[kHZ].

A primary synchronization signal (PSS) and a secondary synchronization signal (SSS) are transmitted in the DL to enable UEs to adjust frequency and timing offsets of the UE. Typical UE devices will incur offsets as a side effect of cheap oscillators, but movement will affect the offsets by varying Doppler and propagation delay.
After synchronization UEs read the master information block (MIB) and system information blocks (SIB)s to obtain vital system parameters. Once synchronised, UEs may connect to the RAN through the RA procedure to perform transmissions.

\begin{center} 
\captionof{table}{Example of static overhead on NB-IoT anchor.}
\begin{tabular}{|lr|lr|}
\hline
\textbf{DL}          & Overhead & \textbf{UL}                & Overhead        \\ \hline
NPSS + NSSS & 15~\%       & PRACH & 28~\%      \\
NRS         & 4~\%        & DMRS              & 10,29~\%   \\
NPBCH       & 9.52~\%     &              &  \\
NB-SIB1     & 4.76~\%     &                   &         \\
NB-SIBx     & 8~\%        &                   &         \\
PDCCH       & 18.15~\%    &                   &         \\
Total       & 59.42~\%  & Total                 & 38,29~\%     \\ \hline
\end{tabular}
\label{tab:overhead}
\end{center}

An example of the static overhead that can be expected on a anchor carrier in a NB-IoT cell is plotted in Tab. \ref{tab:overhead}. Since this overhead is around 60\% in the DL and 40\% in the UL it stands to reason that a non-anchor carrier is added to provide room for dynamic traffic load.

\subsection*{Satellite}
The dimensions of each unit in a CubeSat is $100\times 100\times 100$ [mm], which greatly reduces the amount of surface space available for both solar panels and antennas, but also the internal space available for batteries and RF-subsystems. 

The power budget of a 1U CubeSat for imaging missions allocated enough power to the communication system to achieve $1$~[W] transmission power.\cite{4284088} The orbital average power (OAP) of a 6U Cubesat has been simulated at $17$~[W].\cite{SnyderAndreaniBeerbowerClavijoJoyLeeValeroAraujo2020} Indeed, 100+ watts are obtainable using deployable solar panels.\cite{GOM}
In a LEO satellite intended to be used as a base-station for communication infrastructure greater transmission power is advantageous so we shall assume a satellite of sufficient size to provide a total transmission power of $16$~[W], which results in $8$~[W/Carrier].

The surface-mounted antenna aperture size is limited, so 
a microstrip patch antenna design for $2$ [GHz] has been chosen. The antenna has a patch size of $24\times 33$~[mm] which easily fits on a CubeSat and provides a gain of $8.48$ [dB] for a $3$ [dB] angular width of $73.4$ degrees.\cite{5993433} It should be noted that deployable antennas allow for larger aperture sizes i.e. higher gain within a narrower beam width.

\subsection*{User Equipment}
The UE in this scenario is a standard IoT device based on current TN NB-IoT chip-set. This device has a transmission power of $200$ [mW] (PC3) and a cheap oscillating crystal. The UE must compensate for Doppler to avoid interfering with other UL transmissions. To this end, the base-station node (nodeB) onboard the satellite will broadcast information that allows the position and velocity of the satellite to be estimated. The UE can use this information if it knows it's own position, to calculate and compensate for the Doppler offset and propagation delay. So UEs will either contain a GNSS module or be provisioned with their location if they are stationary.


\newpage
\section*{LINK ANALYSIS}
In this section the communication link in non-terrestrial low-earth orbits is examined. 
First, geometric results are presented for a satellite pass, then results for propagation delay and Doppler frequency offset are presented and finally the link budget and channel fading models are examined.

\subsection*{Geometry}
The geometry of the satellite UE link is depicted in Fig. \ref{fig:satUElink} with key metrics of interest denoted. The curvature of Earth is not shown, but it is taken into account in the model.

\begin{Figure}
    \centering
    \includegraphics[width = 1\textwidth]{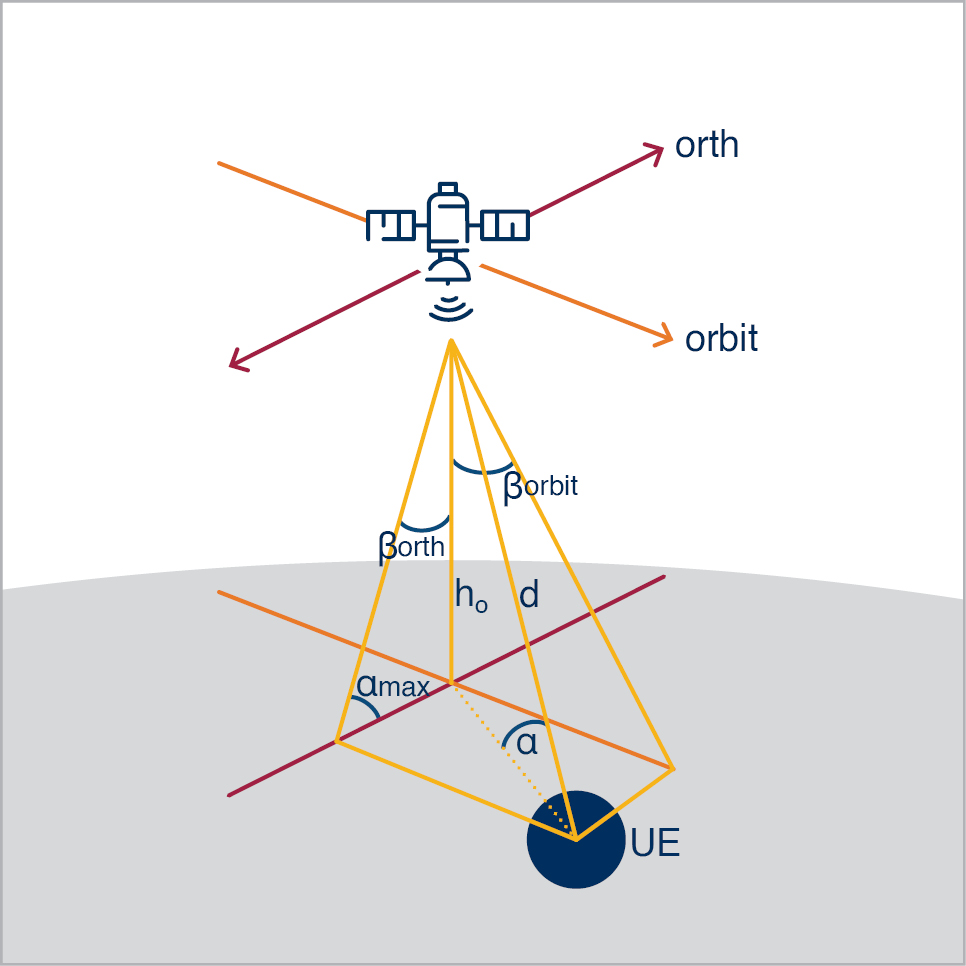}
    \captionof{figure}{UE satellite link geometry.}
    \label{fig:satUElink}
\end{Figure}

The elevation angle, $\alpha$ is the angle between the surface of earth and the direction of the satellite as seen from the UE. 
Let $\alpha_{\min}$ denote the minimum elevation angle at which the satellite is visible to the UE. A typical value for a feeder link would be 10 degrees, whereas 30-40 degrees would be more realistic for IoT devices. 
Here, a satellite pass is the period in which the satellite is within $\alpha_{\min}$ for a UE.
Let $\alpha_{\max}$ be the maximum elevation angle experienced by a UE during a satellite pass. Then $\alpha_{\max}$ fixes the position of a UE in the plane orthogonal to the satellite.

The angle between the UE and nadir as seen by the satellite is denoted $\beta$. This angle can be given in the orbital plane as $\beta^{orbit}$ or the orthogonal (cross-track) plane as $\beta^{orthogonal}$. In the modelled scenario $\beta^{orthogonal}$ is fixed by $\alpha_{\max}$.
The distance between the UE and the satellite is defined as $d$.

\begin{Figure}
    \centering
    \includegraphics[width = 1\textwidth]{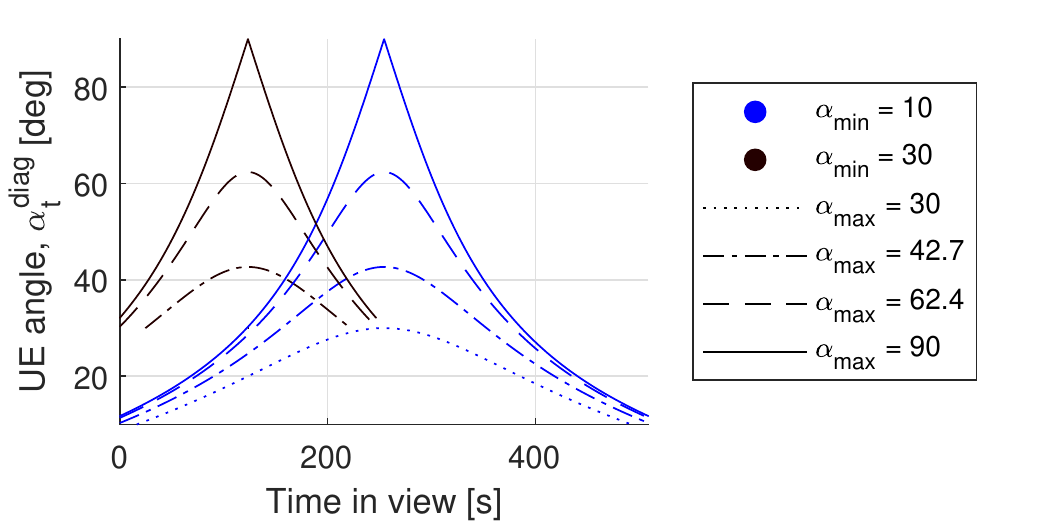}
    \includegraphics[width = 1\textwidth]{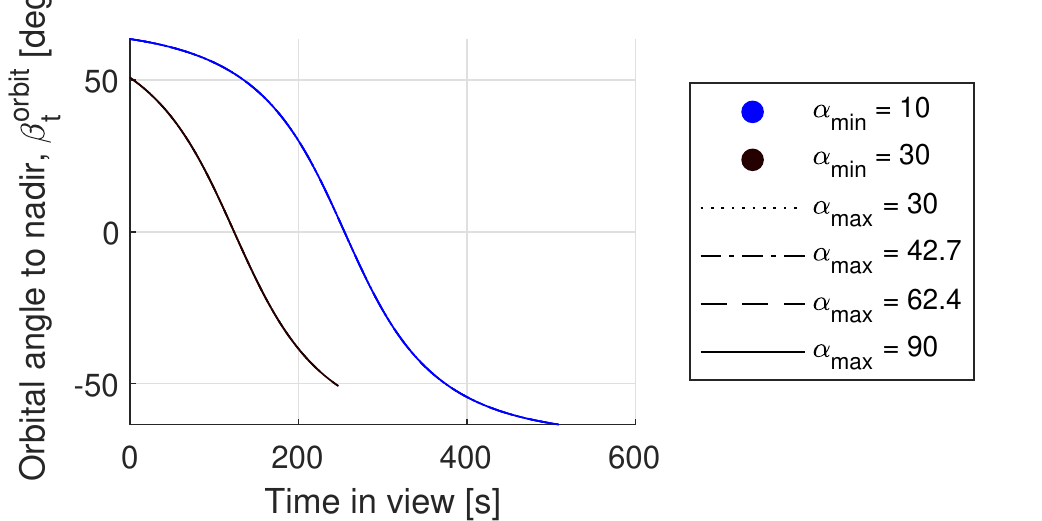}
    \includegraphics[width = 1\textwidth]{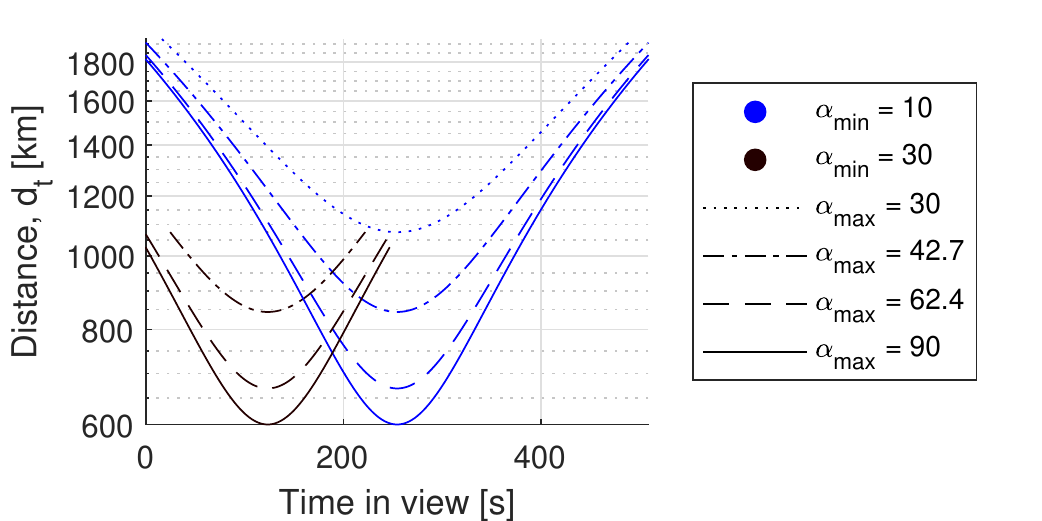}
    \captionof{figure}{Development of key geometric variables during a satellite pass.}
    \label{fig:geoResults}
\end{Figure}

\subsection*{Doppler shift and Propagation delay}
The change in distance over time results in Doppler effect offsetting the frequency of the signals transmitted between the UE and satellite as well as introducing a variable propagation delay between the UE and satellite.\cite{9148880}
The Doppler shift and the propagation times during a satellite pass are plotted in Fig. \ref{fig:Doppler} for S-band communication ($2$ [GHz]). The maximal Doppler offset is $\pm43$ [kHz] when the satellite is far from the UE flying towards the UE with a max rate of change of $544$ [Hz/s] experienced as the satellite passes over the UE.
The propagation delay is less than $4$ [ms] for IoT devices and up to $6.5$ [ms] for feeder links. The maximal delay rate of change is up to $20$ [$\mu$s/s] when the satellite is far from the UE.

\begin{Figure}
    \centering
    \includegraphics[width = 1\textwidth]{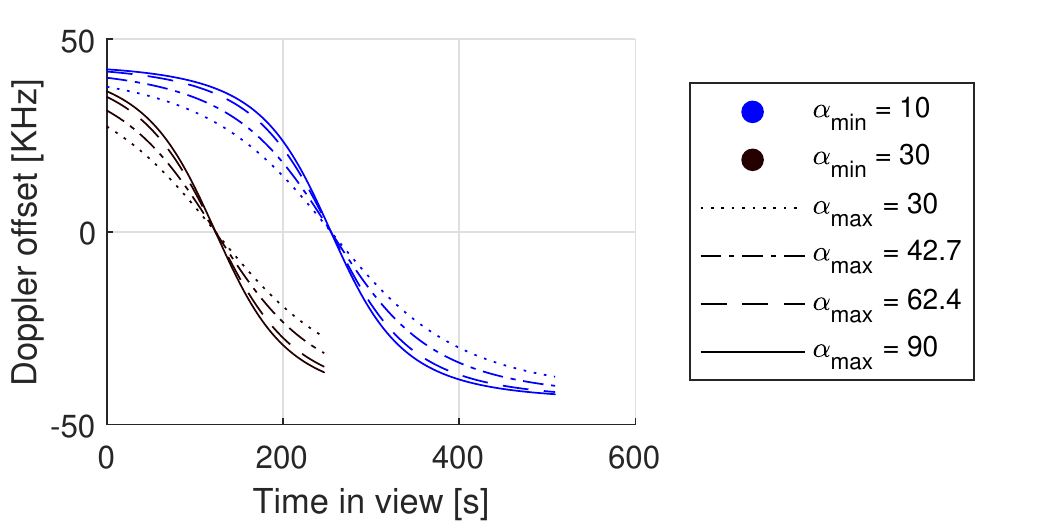}
    \includegraphics[width = 1\textwidth]{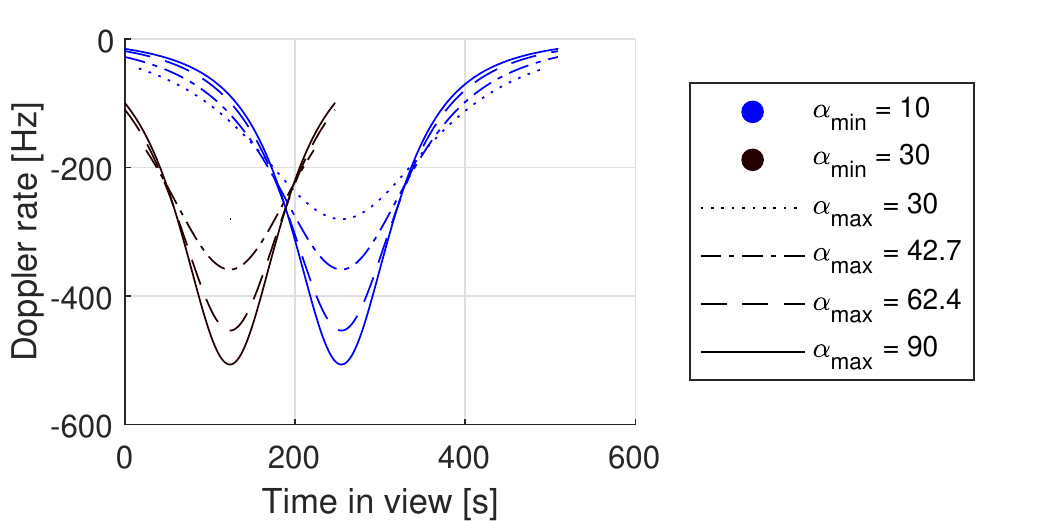}
    \captionof{figure}{Doppler offset (above) and Doppler rate (below) during a satellite pass.}
    \label{fig:Doppler}
\end{Figure}

\begin{Figure}
    \centering
    \includegraphics[width = 1\textwidth]{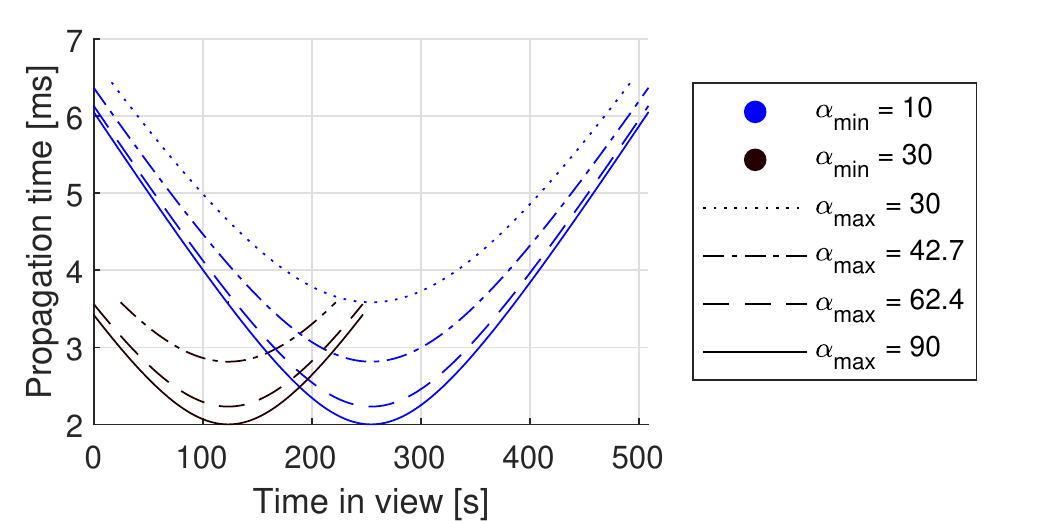}
    \includegraphics[width = 1\textwidth]{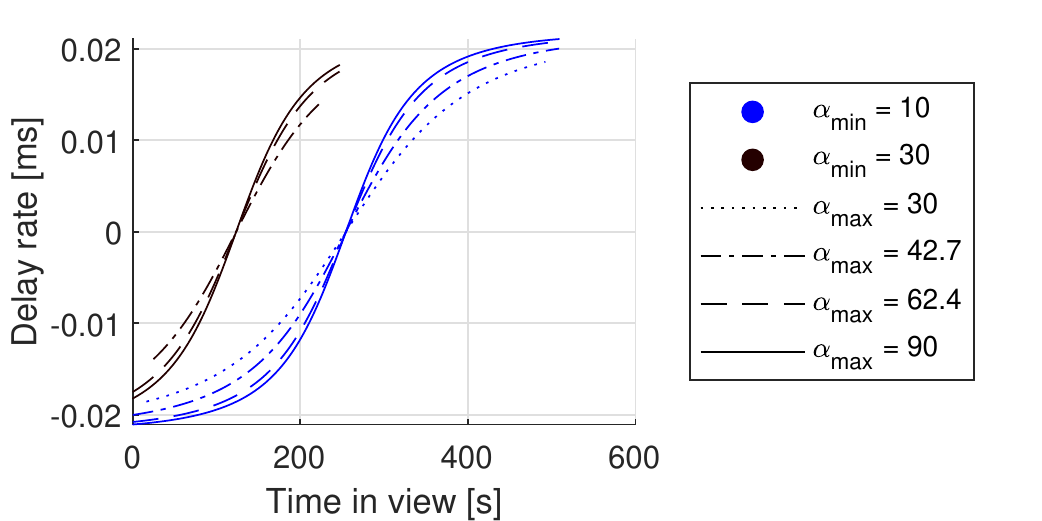}
    \captionof{figure}{Propagation time (above) and delay rate change (below) during a satellite pass.}
    \label{fig:proptime}
\end{Figure}

\subsection*{Link Budget}
The link-budget is the sum of all gains and losses in dB as per Eq.~\eqref{eq:LB}.

\begin{align} \label{eq:LB}
    \textrm{LB} = &P_{TX}+P_N+PL+NF_{RX}+G_{UE,Ant}  \nonumber
\\&+G_{Sat,Ant} + G_{shadow}+G_{polarMiss} \nonumber
\\ &+G_{absorb}+G_{scintillation} 
\end{align}
Where LB is the link-budget (SNR in the receiver) in dB, $P_{TX}$ is the transmission power, $P_N$ is the noise power, PL is the path-loss, $NF_{RX}$ is the receiver's noise-figure, $G_{shadow}$ is the loss due to shadowing, $G_{polarMiss}$ is loss due to polarization mismatch, $G_{absorb}$ is loss due to atmospheric absorption and finally $G_{scintillation}$ is scintillation loss.

\begin{center}
\fontsize{8}{7.2}\selectfont
\captionof{table}{Link budget.}
\begin{tabular}{|l|llllll|}
\hline
&\rotatebox{270}{$\mathrm{NF}_\mathrm{UE}$} &\rotatebox{270}{$\mathrm{NF}_\mathrm{Sat}$}  &\rotatebox{270}{$\mathrm{G}_\mathrm{Shadow}$} &\rotatebox{270}{$\mathrm{G}_\mathrm{polarMiss}$}  &\rotatebox{270}{$\mathrm{G}_\mathrm{absorption}$} &\rotatebox{270}{$\mathrm{G}_\mathrm{scintillation} \;$}
\\ dB & -9 & -3 & -3 & -3 & -0.1 & -2.2 
\\ \hline
\end{tabular}
\label{tab:linkbudget}
\end{center}

Tab. \ref{tab:linkbudget} contains values for the constants of the link budget, Eq.~\eqref{eq:LB}. $G_{absorp}$ and $G_{scintillation}$ depend on the elevation angle but are fixed in this analysis as in the ongoing 3GPP work.\cite{TR36.763} The noise figure of the UE is purposely selected to be quite high and may improve as the technology matures.

The noise power is calculated as thermal noise at the receiver. 
\begin{align} \label{eq:p_n}
    P_N &= 10\log_{10}(k_BT\Delta f)+30 &[dBm]
\end{align}

The path-loss and antenna gains during a satellite pass are a function of the distance and angles between the UE and eNB.

\begin{align} \label{eq:pl}
    PL &= 10\cdot n\cdot\log_{10}(d) &[dBm]
\end{align}

\begin{figure*}
    \centering
    \includegraphics[width = 1\textwidth]{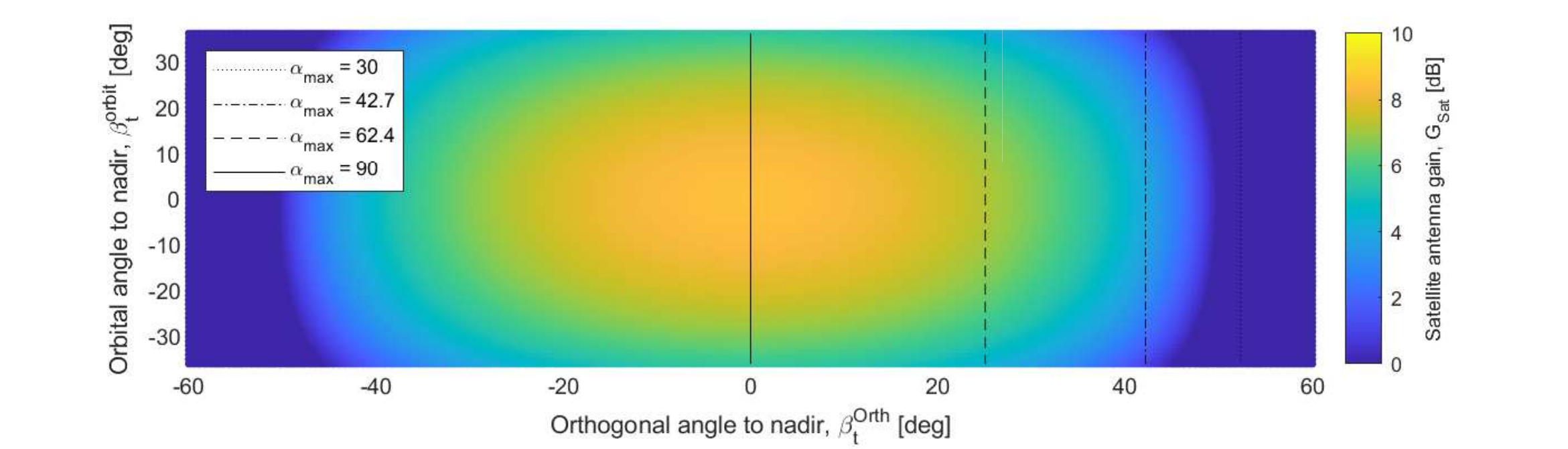}
    \caption{Heat map of satellite antenna gain.}
    \label{fig:satgain}
\end{figure*}

Finally, the gain of the satellite and UE antennas must be accounted for. To simplify the model of the UE, assume that $G_{UE,Ant}=0$  when $\alpha<\alpha_{\min}$. The satellite antenna gain is plotted in Fig. \ref{fig:satgain}.

\begin{figure*}
    \centering
    \includegraphics[width = 1\textwidth, trim=0 25 0 0, clip]{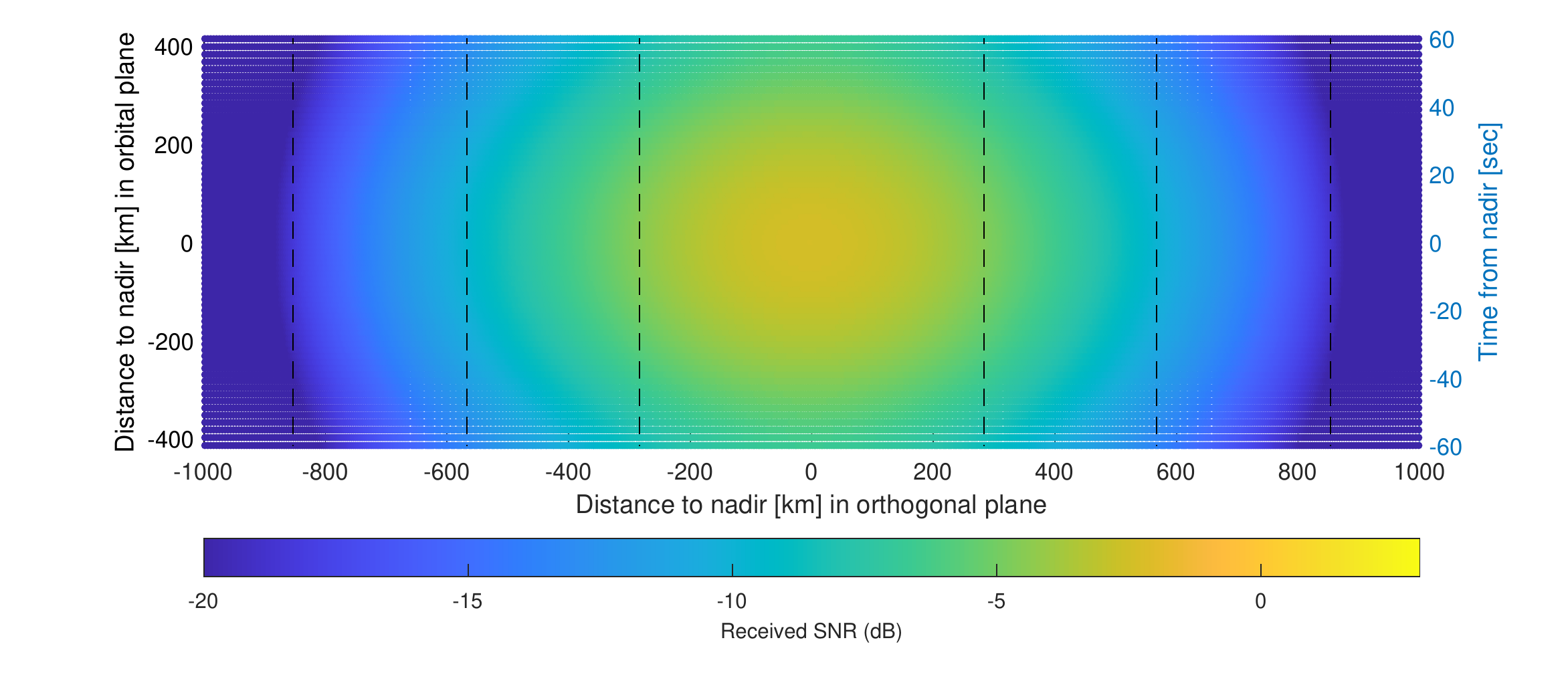}
    \caption{Link-budget in the downlink.}
    \label{fig:dlLB}
\end{figure*}

\begin{figure*}
    \centering
    \includegraphics[width = 1\textwidth, trim=0 25 0 0, clip]{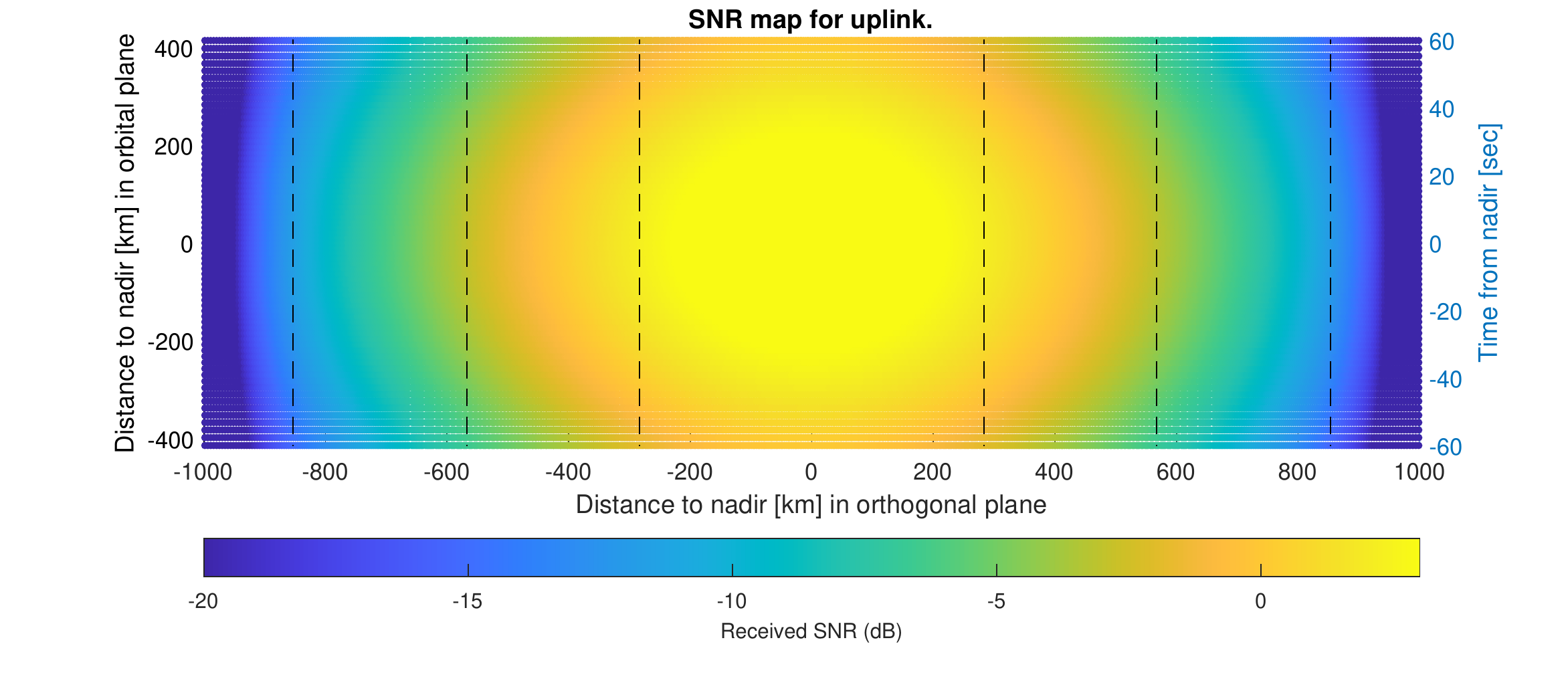}
    \caption{Link-budget in the uplink for 3.75 kHz transmissions.}
    \label{fig:ulLB}
\end{figure*}

The resulting link budget is plotted in Fig.~\ref{fig:dlLB} and Fig.~\ref{fig:ulLB} for the DL and the UL, respectively. The dotted lines correspond to an $\alpha_{\max}$ of $62.4$, $42.7$ and $30$ degrees, respectively.

The DL is power-limited, especially on the beam-edges, while the UL is experiencing much more favorable conditions. The reason for this disparity is the much narrower bandwidth that can be used in the UL, which improves the noise by up to $\sim16.8$~[dB]. In addition, the noise figure in the satellite is a $6$~[dB] improvement over the noise figure in the UEs receiver. Thus, the UL exceeds the $16$~[dB] difference in transmission power by $6.8$~[dB]. 
 
\subsection*{Fading model}
Channel models for NTN were presented in 6.9.2-3 and 6.9.2-4 of 3GPP~TR~38.811.\cite{TR38.811} These fading models are for NTN in urban and hilly environments, respectively. Both models are delay-tab models with a LoS component and Rician gain for all other components. The model parameters are given in Tab.~\ref{tab:fading}. 

\begin{center}
\fontsize{8}{7.2}\selectfont
\captionof{table}{Fading models.}
\begin{tabular}{|l|l|l|}
\hline
Model           & NCU                       & NDH                                 \\ \hline
Environment     & Urban                     & Hilly                               \\
Noise           & AWGN                      & AWGN                                \\
Tab delay [ns] & \{0, 1481\}    & \{0, 168, 2199\}           \\
Tab gain [dB] & \{-10.6, -23.4\}  & \{-11.99, -9.89, -16.77 \} \\
K-factor        & 7                         & 7                                   \\ \hline
\end{tabular}
\label{tab:fading}
\end{center}

\vspace{1cm}

\section*{NB-IOT ADAPTATIONS}
This section presents the main areas in which NB-IoT may need modifications to work in the NTN case and assessments of what modifications, if any, are required.

\subsection*{Detection}
In order to detect a cell a UE must detect the NPSS and NSSS sequences. The time it takes to detect these signals is a function of the SNR. Simulated results for this synchronisation time are given in Tab. \ref{tab:cellsearch}.

\begin{center} 
\captionof{table}{Required number of frames for cell search success.}
\resizebox{\columnwidth}{!}{\small
\begin{tabular}{lrccc|}
\cline{3-5}
                                          & \multicolumn{1}{l|}{}             & \multicolumn{3}{l|}{\textbf{Doppler}}                                                                                                                                                                                                          \\ \hline
\multicolumn{1}{|l|}{\textbf{Model}}      & \multicolumn{1}{l|}{\textbf{SNR}} & \textbf{\begin{tabular}[c]{@{}c@{}}28.4 {[}kHz{]} \\ 306 {[}Hz/s{]}\end{tabular}} & \textbf{\begin{tabular}[c]{@{}c@{}}0 {[}Hz{]} \\ 580 {[}Hz/s{]}\end{tabular}} & \textbf{\begin{tabular}[c]{@{}c@{}}0 {[}Hz{]}\\ 0 {[}Hz/s{]}\end{tabular}} \\ \hline
\multicolumn{1}{|l}{\multirow{5}{*}{LOS}} & -10~dB                             & 532                                                                               & 414                                                                           & 426                                                                        \\
\multicolumn{1}{|l}{}                     & -7~dB                              & 30                                                                                & 26                                                                            & 28                                                                         \\
\multicolumn{1}{|l}{}                     & -4~dB                              & 4                                                                                 & 4                                                                             & 4                                                                          \\
\multicolumn{1}{|l}{}                     & 0~dB                               & 2                                                                                 & 2                                                                             & 2                                                                          \\
\multicolumn{1}{|l}{}                     & +5~dB                              & 2                                                                                 & 2                                                                             & 2                                                                          \\ \hline
\multicolumn{1}{|l}{\multirow{5}{*}{NCU}} & -10~dB                             & 3110                                                                              & 3350                                                                          & 2450                                                                       \\
\multicolumn{1}{|l}{}                     & -7~dB                             & 64                                                                                & 60                                                                            & 50                                                                         \\
\multicolumn{1}{|l}{}                     & -4~dB                              & 10                                                                                & 8                                                                             & 8                                                                          \\
\multicolumn{1}{|l}{}                     & 0~dB                              & 2                                                                                 & 2                                                                             & 2                                                                          \\
\multicolumn{1}{|l}{}                     & +5~dB                             & 2                                                                                 & 2                                                                             & 2                                                                          \\ \hline
\multicolumn{1}{|l}{\multirow{5}{*}{NDH}} & -10~dB                             & 586                                                                               & 490                                                                           & 436                                                                        \\
\multicolumn{1}{|l}{}                     & -7~dB                             & 42                                                                                & 40                                                                            & 34                                                                         \\
\multicolumn{1}{|l}{}                     & -4~dB                              & 8                                                                                 & 8                                                                             & 8                                                                          \\
\multicolumn{1}{|l}{}                     & 0~dB                               & 4                                                                                 & 4                                                                             & 4                                                                          \\
\multicolumn{1}{|l}{}                     & +5~dB                              & 2                                                                                 & 2                                                                             & 2                                                                          \\ \hline
\end{tabular}
}
\label{tab:cellsearch}
\end{center}

The periods in which UEs may potentially have detected NPSS/SSS during a pass are plotted in Fig. \ref{fig:sync}. These periods are given for a UE that is searching all the time, but in practise a UE would attempt cell search at a certain time, which is not necessarily the moment it enters the cell edge of a satellite pass.

\begin{Figure}
    \centering
    \includegraphics[width = 1\textwidth]{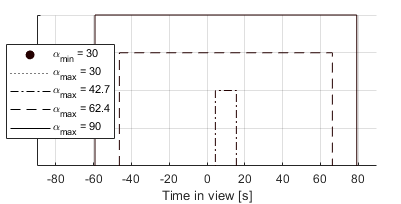}
    \captionof{figure}{Synchronization windows during a satellite pass.}
    \label{fig:sync}
\end{Figure}

\begin{figure*}
    \centering
    \includegraphics[width = 1\textwidth]{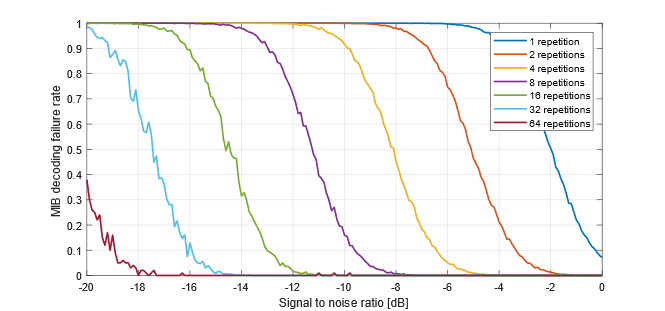}
    \caption{MIB decoding performance.}
    \label{fig:mib}
\end{figure*}

The periods above do not account for decoding the MIB and SIB, which is not a negligible amount of time. The number of repetitions required to decode a MIB successfully as a function of the received SNR is plotted in Fig. \ref{fig:mib}. Evidently, it is only UEs for which $\alpha_{\max} \leq 42.7$ [degrees], or equivalently within a surface distance of $283$~[km] of nadir, that are able to access the cell for the considered link budget.

\subsection*{Synchronization}
In order to avoid interference in the UL UEs must be able to synchronize transmissions in time and frequency a-priori to their reception at the eNB. 

To facilitate this it has been decided in 3GPP that satellites should broadcast information that allows for the position and velocity of the satellite to be computed either in direct form or indirectly as ephemeris information. UEs must also know their own position either through provisioning for stationary devices, or through GNSS measurements. Given both the UE's and satellite's position and velocity are known to the UE it can calculate both current and future Doppler offset and propagation delay.

The first transmission from a UE will be a random access preamble (RAP) in the random access channel (RACH). The eNB will respond with a random access response (RAR) message, which will include timing advance (TA) and frequency advance (FA) components letting the UE know how far off the expected time and frequency the RAP was. The UE will use this information to tune the time-frequency adjustment.

During long UL transmissions the drift in frequency and propagation delay may be excessive if the UE does not adjust. Here the UE could compensate discretely over sequences of the transmission, which would require the transmission to be broken up and pauses to be inserted between each part to allow for timing adjustments. The pauses could even be used by the eNB to provide TA adjustments, but this has potential for creating large amounts of overhead in the DL. Another option is to process the output RF signal with a time-dependent continuous filter that is the inverse of the predicted Doppler offset and propagation delay.

\subsection*{PHY rate}
The required SNR for a block error rate of 10\% for various modulation and coding schemes (MCS) has been found through simulation\cite{10.1007/978-3-030-31831-4_18}, extrapolated for maximum ratio combining (MRC) and is given in Tab.~\ref{tab:DLtbssnr}. The achievable PHY-rate in the DL for a 100 bit payload during a satellite pass is plotted in Fig. \ref{fig:phyrate}.

\begin{Figure}
    \centering
    \includegraphics[width = 1\textwidth]{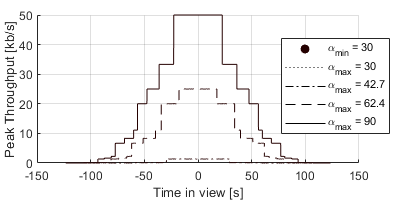}
    \captionof{figure}{Potential DL PHY-rate during a satellite pass.}
    \label{fig:phyrate}
\end{Figure}
\vspace{1 cm}
\begin{center}
\captionof{table}{Simulated SNR requirement for various modulation and coding schemes (MCS).}
\begin{tabular}{|c|rrrrr|} \hline
& \multicolumn{5}{l|}{Repetitions}   \\
I\_TBS & 1    & 2    & 4    & 8    & 16    \\ \hline
0& -5,8 & -8,3 & -10,6 & -12,8 & -14,7 \\
1& -4,9 & -7,2 & -9,7  & -11,9 & -13,8 \\
2& -3,9 & -6,2 & -8,8  & -11   & -12,9 \\
3& -3   & -5,4 & -8    & -10,4 & -12,2 \\
4& -2   & -4,6 & -7,2  & -9,6  & -11,4 \\
5& -1,1 & -3,7 & -6,3  & -8,9  & -10,8 \\
6& -0,2 & -2,8 & -5,6  & -8    & -10   \\
7& 0,7  & -1,9 & -4,7  & -7,3  & -9,3  \\
8& 1,4  & -1,3 & -4,1  & -6,8  & -8,9  \\
9& 2,2  & -0,4 & -3,3  & -6    & -8,1  \\
10& 3,1  & 0,4  & -2,4  & -5,2  & -7,3  \\
11& 4,2  & 1,4  & -1,5  & -4,3  & -6,6  \\
12& 5,5  & 2,7  & -0,4  & -3,3  & -5,6  \\
13& 6,9  & 3,9  & 0,9   & -2    & -4,4 \\ \hline
\end{tabular}
\label{tab:DLtbssnr}
\end{center}

\subsection*{Random Access}
As previously noted the random access preamble is the first transmission a UE makes in the UL. The requirements for this initial transmission is a frequency offset within $\pm 200$~[Hz]\cite{TR38.101} and a max timing advance that is by Eq. \eqref{eq:TA}.

\begin{align} \label{eq:TA}
TA_{\max} = \pm\dfrac{T_{CP}}{4} = \begin{cases}
        \pm16.75~\mu s \;,\;\text{for format 0} 
        \\
        \pm66.75~\mu s \;,\;\text{for format 1} 
        \end{cases}   
\end{align}

The SNR of the received preambles will determine how many preambles are necessary for accurate detection, timing- and frequency offset estimation at the eNB. The simulation results given in Tab. \ref{tab:rap} in tandem with the UL link-budget of Fig. \ref{fig:ulLB} indicate that the RACH may not need to be configured towards coverage extension (CE) levels that take up more spectrum as a relatively low number of repetitions are sufficient.

\begin{center}
\captionof{table}{Repetition requirements for simulated RAP reception. Detection failure percentage in parenthesis.}
\begin{tabular}{@{}|l|lll|@{}}
\hline
SNR & AWGN      & NCU & NDH        \\ \hline
0                         & 2         & 1   & 2         \\
-4                        & 8        & 4  & 8        \\
-7                        & 32        & 8  & 32  \\
-10                       & 128       & 62  & 128 \\
-12                       & 128 (8\%) & 128 & 128 (13\%) \\ \hline
\end{tabular}
\label{tab:rap}
\end{center}

\subsection*{Scheduling}
Timers should be offset by an integer $T_k$ [ms] to accommodate the propagation delay such that $T_k~>~d/C$.
The minimum viable solution here is to set $T_k$ to a fixed value and transmit it in the MIB or SIB. Alternatively, it could be the responsibility of the UE to compute $T_k$ over time and signal to the eNB when $T_k$ changes by a millisecond.

\subsection*{Paging occasions}
The core network and UEs must synchronize and agree upon paging opportunities. The alternative is that UEs observe the downlink control channel all of the time in order to not miss any paging, which is of course incredibly costly in terms of energy, or that the UE unwittingly ignores paging opportunities and so becomes unreachable by the core network.

Conventionally, discontinuous reception (DRX) and Power Save Mode (PSM) have been used to save energy by creating a formal agreement on paging ocassions between the core network and UEs and letting each UE go into a power saving mode outside of the paging ocassion. In LEO NTN NB-IoT this is complicated by the movement of cells, especially in low density constellations, where UEs may be out of coverage for extended periods of time.

A minimum viable way to accommodate this situation is to utilize PSM for UEs whilst out of coverage and then change to iDRX mode when within coverage of a satellite.


\subsection*{Discontinuous feeder links}
In constellations without continuous feeder links transparent payloads are per definition blocked. Given the relatively small range of LEO satellites and the small coverage window for UEs the transparent payload scenario is likely to perform poorly for LEO NTN~NB-IoT in general. Cross-ocean availability of feeder-links is for example not expected for LEO satellites.

In the case of regenerative payloads one way to support feeder link discontinuity is to keep part of the core network entities, such as the mobility management entity (MME) on-board the satellite. This could allow UEs to access the RAN and perform transmissions or allow for paging in the DL even when the satellite is our of range of a feeder link. In combination with a store-and-forward scheme this could allow support of devices in areas far from ground-stations.

\section*{CONCLUDING REMARKS}

\subsection*{Discussion}
The UL link-budget is $6.8$~[dB] greater than the DL link-budget. This also means that a total increase of $60$~[W] of the DL power budget will equalize the link budget for DL and $3.75$ [kHz] UL transmissions. Such a power budget may not be feasible for nano-satellites, but it should be within range for 'larger' small-satellites.

The limited DL link-budget may not be too much of a draw-back in the IoT scenario with regards to DL capacity as many IoT applications favor the UL, but a more direct consequence is the limited area in which synchronization to the cell is possible before it moves out of range. The power budget and antenna configuration presented here allow for DL synchronization within a relatively small cell ($500$ [km]). The cell size could be increased by changing the antenna configuration and upping the power budget.

Another trade-off beyond capacity and coverage of single satellites, is the trade-off between satellite cost, coverage/capacity and constellation density.

\subsection*{Conclusion}
This paper has presented the theoretical and simulated results of a link level analysis of LEO NTN~NB-IoT. It was shown that a nano-satellite could serve as a functioning base-station in a NTN~NB-IoT cellular infrastructure albeit at a diminished cell size of $560$~[km] across. Furthermore enhancements related to low-density constellations and discontinuous feeder links were discussed.

\newpage
\subsection*{Further work}
GateHouse will continue to develop our NTN NB-IoT waveform and is planning an in-orbit demonstration to validate the link-budget, adaptations and algorithms. Furthermore, the standardization activities in 3GPP for 5G NTN will continue with active participation from GateHouse.

\subsection*{Acknowledgments}
Thanks to our colleagues at GateHouse, in particular Bertel~Brander, Gert~Børsen, Johannes~Elgaard and Lars~Weje~Hangstrup for their work on the PHY layer algorithms, simulations and implementation of the NTN~NB-IoT waveform.

The work documented in this paper was in part financed by ESA 5997 NB-IoT: 
'Narrowband IoT standard for Smallsat Networks'.

\bibliography{bib.bib}
\bibliographystyle{unsrt}

\end{multicols*}
\end{document}